\documentclass[10pt,journal,compsoc]{IEEEtran}
\usepackage{dcolumn}
\usepackage{bm}
\usepackage{amsmath}
\usepackage{amssymb,amsthm}
\usepackage{bbm}
\usepackage{epsfig,color}
\usepackage[sans]{dsfont}
\usepackage{comment}
\usepackage{hyperref}
\usepackage{units}
\usepackage{color}
\usepackage{ifthen}
\usepackage{graphicx}
\graphicspath{ {images/} }

\usepackage{mathrsfs}

\usepackage[utf8]{inputenc}
\usepackage{tikz}
\usetikzlibrary{automata,positioning}
\usetikzlibrary{fit,shapes.geometric}
\usetikzlibrary{decorations.pathmorphing}
\usetikzlibrary{arrows,matrix}

\usepackage{mathtools}

\definecolor{nblue}{rgb}{0.2,0.2,0.7}
\definecolor{ngreen}{rgb}{0.2,0.6,0.2}
\definecolor{nred}{rgb}{0.7,0.2,0.2}
\definecolor{nblack}{rgb}{0,0,0}

\newcommand{\tr}{\text{tr}}

\renewcommand{\H}{\mathcal{H}}

\def\B{\mathcal{B}}

\newcommand{\N}{\mathcal{N}}

\newcommand{\x}{\textrm{x}}

\newtheorem*{theorem*}{Theorem}

\theoremstyle{definition}

\theoremstyle{plain}

\theoremstyle{plain}

\theoremstyle{plain}

\theoremstyle{plain}

\theoremstyle{plain}

\providecommand{\conjecturename}{Conjecture}
\providecommand{\definitionname}{Definition}
\providecommand{\lemmaname}{Lemma}
\providecommand{\corollaryname}{Corollary}
\providecommand{\theoremname}{Theorem}
\providecommand{\propositionname}{Proposition}

\def\x{\mathrm{x}}
\def\y{\mathrm{y}}

\def\i{\mathrm{id}}

\def\N{\mathcal{N}}

\def\H{\mathcal{H}}

\def\id{\mathbb{I}}

\def\tr{\mbox{tr}}

\def\bea{\begin{eqnarray}}
\def\eea{\end{eqnarray}}

\ifCLASSOPTIONcompsoc
  \usepackage[nocompress]{cite}
\else
  \usepackage{cite}
\fi
\ifCLASSINFOpdf
\else
\fi
\begin{document}
%

\title{ A hybrid quantum-classical approach to mitigating measurement errors }

%
%
%
%

\author{Hyeokjea~Kwon$^{*}$~
        and~Joonwoo~Bae$^{\dagger}$
\IEEEcompsocitemizethanks{\IEEEcompsocthanksitem H. Kwon and J. Bae are with the School of Electrical Engineering, Korea Advanced Institute of Science and Technology, Daejeon, Republic of (South) Korea.\protect\\
E-mail: $^{*}hyukjaekwon@kaist.ac.kr$, $^{\dagger}$joonwoo.bae@kaist.ac.kr
} 
}

%
%

\markboth{}
{Kwon \MakeLowercase{\textit{et al.}}: Bare Demo of IEEEtran.cls for Computer Society Journals}
%



\IEEEtitleabstractindextext{%
\begin{abstract}
When noisy intermediate scalable quantum (NISQ) devices are applied in information processing, all of the stages through preparation, manipulation, and measurement of multipartite qubit states contain various types of noise that are generally hard to be verified in practice. In this work, we present a scheme to deal with unknown quantum noise and show that it can be used to mitigate errors in measurement readout with NISQ devices. Quantum detector tomography that identifies a type of noise in a measurement can be circumvented. The scheme applies single-qubit operations only, that are with relatively higher precision than measurement readout or two-qubit gates. A classical post-processing is then performed with measurement outcomes. The scheme is implemented in quantum algorithms with NISQ devices: the Bernstein-Vazirani algorithm and a quantum amplitude estimation algorithm in $\mathrm{IBMQ\underbar{~}yorktown}$ and $\mathrm{IBMQ\underbar{~}essex}$. The enhancement in the statistics of the measurement outcomes is presented for both of the algorithms with NISQ devices.
\end{abstract}

}

\maketitle

\IEEEdisplaynontitleabstractindextext

%
\IEEEpeerreviewmaketitle


%
%
%
%

\section{Introduction}
The state of the art quantum technologies can be represented by noisy intermediate scalable quantum (NISQ) devices, by which multiple qubits can be prepared, connected, controlled, extended, and also networked but, all of them are essentially contain noise. Quantum algorithms having advantages over their classical counterparts, such as quantum period-finding \cite{shor, simon}, quantum search \cite{grover}, quantum factoring \cite{shor}, and quantum oracle algorithms \cite{deutsch, bv, ambainis}, etc., can be realized by the building blocks, quantum Fourier transform (QFT) \cite{book} and quantum amplitude amplification (QAA) \cite{grover, qaa}. To this end, it is crucial to apply coherent quantum operations on multiple qubits as the building blocks are constructed by entangling gates on multiple qubits sequentially. When the algorithms are realized with NISQ devices, noise that essentially appears in all of the stages from preparation to measurement readout are not controlled and thus accumulated. Consequently, no quantum advantage may be achieved.

Then, one of the key problems is to find how useful the NISQ devices are, possibly, for practical purposes \cite{preskill}. Namely, it is sought to have quantum advantages beyond the limitations of today's technologies even though the devices do not ideally work as expected as true quantum devices, e.g., \cite{google, konig}. Recently, quantum algorithms that are better fitted with NISQ devices have been devised. For instance, variational quantum eigensolvers \cite{vqe} and quantum approximate optimization algorithms \cite{qaoa} have been proposed, which are designed in such a way that errors during quantum gates are not accumulated. Note that with NISQ devices, quantum error correction which preserves the purity of quantum states through all the process is lacking \cite{qec}. 

In the other way around, there have been attempts to scrutinize NISQ devices applied to performing quantum algorithms. For instance, in the case of IBM Q devices it is found that single-qubit gates are with relatively high precisions with an error rate about $0.1\%$ while two-qubit gates and measurement readout are comparable with errors about $2$-$10\%$ \cite{data} \cite{ion}. The possibility of crosstalk in multiple detectors in measurement readout has been pointed out \cite{crosstalk}. While observing the numbers, one can naturally suggest that single-qubit gates may be used mitigate errors in NISQ devices as their performance is significantly finer than the others. Along the line, there has been extensive effort to develop methods of error mitigation in NISQ information processing, see e.g., \cite{allnisq, n1, n2, osm, gtech}. 

In this work, we present a quantum-classical hybrid method of mitigating errors in measurement readout. The method consists of quantum pre-processing with single-qubit gates only, which are applied right before a detection event, and classical post-processing, which is performed with measurement outcomes. The quantum pre-processing corresponds to the channel-twirl with few single-qubit gates by which the type of errors that are dealt in the classical post-processing can be fixed as those originated from a depolarization map. The scheme of measurement error mitigation is applied to the Bernstein-Vazirani (BV) algorithm based on oracle queries, and the quantum amplitude estimation (QAE) algorithm that contains both QFT and QAA. It is shown that the statistics of measurement outcomes in the algorithms can be enhanced by the proposed scheme. With IBM Q devices that realize $3$-qubit BV and QAE algorithms, it is also shown that measurement readout can be improved. The proposed scheme may be useful in the microarchitecture of NISQ information processing.


The paper is structured as follows. In Sec. \ref{theory}, we explore a model of measurement readout and devise the method of mitigating measurement errors. The assumptions under which the method is effective are addressed. In Sec. \ref{demonstration}, the scheme is applied to the BV and QAE algorithms. It is demonstrated with a Qiskit simulator that the statistics of measurement outcomes can be enhanced by the proposed scheme. In Sec. \ref{IBM}, the scheme is patched on measurement readout in IBM quantum devices, $\mathrm{IBM\underbar{~}yorktown }$ for the QAE algorithm and $\mathrm{IBM\underbar{~}essex }$ for the BV algorithm. The results that show enhancement in the statistics of measurement outcomes is shown. In Sec. \ref{remark}, we discuss future directions and conclude the results.

\section{ Proposal }
\label{theory}

Let us begin with the notations and terminologies to be used throughout. Quantum states are generally described by unit-trace and non-negative operators, denoted by $\rho$, on a Hilbert space $\H$. The set of quantum states is denoted by 
\bea
S(\H) = \{ \rho \in \B(\H)~:~\rho \geq 0,~\tr [ \rho] = 1 \} \nonumber
\eea
where $\B (\H)$ denotes a set of bound linear operators. Qubit states are characterized in a two-dimensional Hilbert space, $\H_2 = \mathrm{span}\{|0\rangle, |1\rangle \}$. Note that in a quantum algorithm an initial state is prepared in $|0\rangle^{\otimes n}$. A measurement is performed in the computational basis,
\bea
M_0 = |0\rangle\langle0|,~~ M_1 = |1 \rangle \langle 1| \label{eq:comm}
\eea
which is a complete measurement, i.e., $M_0 +M_1 =\id$. The postulate of quantum theory states that when a system is prepared in a state $\rho_{\y}$, a measurement outcome $\x$ is obtained with with a probability $\mathrm{P[\x | \y] = \tr[M_{\x} \rho_{\y}] }$. Note that a general measurement is described by positive-operator-valued-measures (POVMs), $\{M_{i}\geq0 \}$ such that $\sum_{i}M_i =\id$.

A quantum dynamics corresponds to a unitary transformation of quantum states, $\rho \mapsto U\rho U^{\dagger}$ where $U^{\dagger}U =UU^{\dagger}=\id$. The operator basis for qubit states is given by the set of Pauli matrices, 
\bea
 \mathbb{I},~X =  \left[ \begin{array}{ccc} 0 &  1  \\ 1  & 0  \end{array}  \right] ,~Y =   \left[ \begin{array}{ccc} 0 & -  i \\  i  & 0  \end{array} \right],~ \mathrm{and}~Z = \left[ \begin{array}{ccc} 1 & 0 \\ 0  & -1  \end{array}  \right]. \nonumber 
\eea
Quantum gates that process dynamics of qubit states are described by unitary transformations. A single-qubit rotation in a Bloch sphere can be described by a unitary transformation in the following, 
\bea
U_{\hat{n}} (\theta)=\exp[ -i\theta \vec{n} \cdot \vec{ \sigma } / 2 ] = \sum_{k=0}^{\infty} \frac{(-i\theta)^k}{ k ! } (\vec{n} \cdot \vec{ \sigma } / 2 )^k  \nonumber
\eea 
where $\sigma= (X,Y,Z)$, meaning a counterclockwise rotation by the angle $\theta$ about the axis $\hat{n}$. For instance, 
\bea
R_y (\theta) = \left[ \begin{array}{ccc} \cos (\theta/2)& - \sin (\theta/2) \\ \sin (\theta/2) & \cos (\theta/2) \end{array} \right] \label{eq:ry}
\eea
shows a counterclockwise rotation on the $x$-$z$ plane by $\theta$ about the $y$-axis. 

For two-qubit states, a controlled-$U$ gate denoted by $C(U)$ with a single-qubit gate $U$ is written as
\bea
C(U) = |0\rangle_c \langle 0|\otimes \id_t + |1\rangle_c \langle 1|\otimes U_t \nonumber
\eea
where $c$ denotes a control qubit and $t$ target qubit. The controlled-NOT (CNOT) gate can be written as, $U_{\mathrm{CNOT}} = C(X)$ with the Pauli matrix $X$. Then, a quantum algorithm can be efficiently composed by a sequence of single- and two-qubit gates only in general \cite{solovay}. 

Quantum circuits that realize quantum algorithms contain both system and ancilla qubits. While the overall dynamics of both system and ancilla qubits is described by a unitary transformation, the reduced dynamics to system ones can be characterized by a positive, completely positive, and trace-preserving map, called a quantum channel, $\Lambda : \rho \mapsto \Lambda[\rho]$, i.e. $\i_k \otimes \Lambda \geq 0$ for all $k\geq1$ with $\i_k$ an identity map over a $k$-dimensional Hilbert space and $\tr[A] = \tr[\Lambda[A]]$ for all $A\in\B(\H)$.

\begin{figure}[h]
\begin{center}	
	\includegraphics[width=0.45\textwidth]{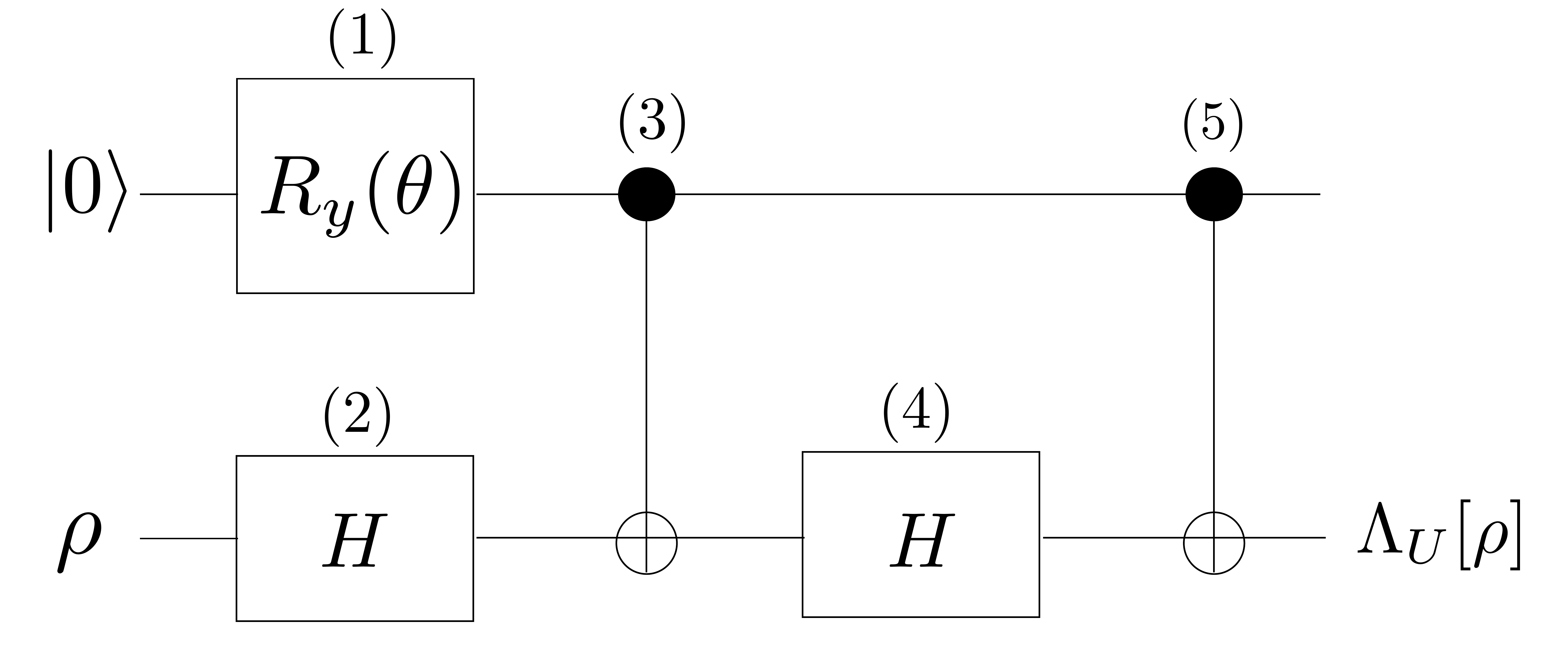}
	\caption{ Quantum circuits that prepare a channel $\Lambda_U [ \cdot ] = (1- p ) \i (\cdot ) + p U (\cdot ) U^{\dagger}$ for $U=X,Y,Z$ are shown. An ancilla qubit is prepared in a state $|0\rangle$ and the operation $R_y (\theta)$ is in Eq. (\ref{eq:ry}). Note that the parameters are related as $p=\sin^2 (\theta/2)$. The Hadamard gate is denoted by $H$. A channel $\Lambda_X$ is constructed by components $(1)$ and $(3)$, $\Lambda_Y$ with $(1)$, $(2)$, $(3)$, $(4)$, and $(5)$ and $\Lambda_Z$ with $(1)$, $(2)$, $(3)$, and $(4)$. For instance, $\Lambda_Z[\rho] = \tr_{\mathrm{anc}} [  (\id \otimes H) U_{\mathrm{CNOT}}  (R_y(\theta)\otimes H )(|0\rangle_{\mathrm{anc}}\langle0| \otimes \rho)  (R_{y}^{\dagger}(\theta)\otimes H)  U_{\mathrm{CNOT}} (\id \otimes H) ]  $. The circuit is exploited to implement noise in a measurement, see Figs. \ref{BVsimulation} and \ref{fig:qae}. } 
	\label{noisemodel}
\end{center}	
\end{figure}

\subsection{ Measurement readout}

Let $U_{\mathrm{circuit}}$ denote a unitary transformation designed to execute a quantum algorithm. The resulting state right before a measurement can be written as,
\bea
|\varphi  \rangle_{1\cdots n,A} = U_{\mathrm{circuit}} ~(|0\rangle^{\otimes n} \otimes |A\rangle) \label{eq:nstate}
\eea 
where $A$ collects all of the ancilla qubits. For $n$ system qubits, a POVM element that gives rise an outcome $\vec{\x} = \x_1 \cdots \x_n$ can be written as 
\bea
M_{\vec{\x}}=M_{\x_1} \otimes \cdots \otimes M_{\x_n},~ \mathrm{where}~ \x_k \in \{ 0,1\}~\forall  k=1,\cdots,n,\nonumber
\eea
which takes place with probability $\tr[M_{\vec{\x}} |\varphi\rangle_{1\cdots n,A} \langle \varphi| ]$. The description of a quantum measurement is consistent to detection events in a realistic scenario, e.g. quantum experiments.

Then, suppose that errors appear in a measurement or observed in the statistics of measurement outcomes. It is, however, not clear how to find the source of errors since a measurement is from interactions of a state and a detector, i.e., a POVM element. A state before a detector may contain noise in advance, or a detector itself is noisy. In what follows, we present a framework that makes distinction between two sources of noise, either a state or a detector. We then find the operational meaning of a measurement in quantum algorithms as optimal state discrimination. Collecting the results, we propose a scheme of error mitigation.

\subsubsection*{Characterizing the sources of noise}

When a measurement is performed on the $k$-th qubit in Eq. (\ref{eq:nstate}), the probability of obtaining a measurement outcome $\x_k$ is given as follows, 
\bea
\mathrm{P} [ \x_k| k]  = \tr[M_{\x_k} \rho^{(k)}],  \label{eq:dec}
\eea
where $\rho^{(k)} = \tr_{\bar{k}} |\varphi\rangle_{1\cdots n,A} \langle \varphi|$ and $\tr_{\bar{k}}$ denotes the partial trace all qubits but the $k$-th one. 

By noise in a detection event, the probability in Eq. (\ref{eq:dec}) may be corrupted. This is often referred to as noise on a POVM element, i.e., $M_{\x_k} \mapsto \N [M_{\x_k} ]$ for a trace-preserving and completely positive map $\N$. After all, the probability is corrupted as, $\tr[ \N [ M_{\x_k} ] \rho^{(k)}]$, which can be estimated in an experiment. We, however, note that the same probability can also be obtained equivalently by noise on a state, 
\bea
\tr[ \N [ M_{\x_k} ] \rho^{(k)}] = \tr[  M_{\x_k}   \N^{\dagger} [ \rho^{(k)}] ] \label{eq:dual}
\eea   
for a dual map $\N^{\dagger}$. From the statistics of measurement outcomes, one cannot find which one, the state or a detector, is noisy.

In the view of observing the statistics of measurement outcomes only, let us now characterize what types of noise can make contributions to the errors in measurement readout. On the one hand, a resulting qubit after the last gate in a circuit and then placed right before a measurement may have experienced unwanted interactions with an environment, i.e., noise may be contained. This can be described by a noisy channel on a qubit state $\Lambda: \rho \mapsto \Lambda[\rho]$. Although this type of noise is not related to a detector, it happens that the source of noise contributes to errors in measurement readout. For instance, as it is shown in Eq. (\ref{eq:dual}), one cannot identify which of a state or a detector is noisy from errors in measurement readout. Moreover, when a quantum algorithm is structured by state preparation, a quantum circuit, and measurement readout, noise on a qubit state between the circuit and a detector contributes to the errors in measurement readout. Therefore, noise on a qubit state before a detection event, more precisely between the last gate of a circuit and a detector, should be taken into account when one attempts to mitigate errors in measurement readout. On the other hand, a detector {\it per se} may not work as desired, which can also be described by a trace-preserving and completely positive map $\N$ on a POVM element, $M_{\x_k} \mapsto \N[M_{\x_k}]$. 

Overall, the probability of a measurement outcome containing readout errors can be characterized by noise on a state and a detector both as follows,
\bea
\mathrm{P}_{ (\N , \Lambda)}[ \x_k | k ] = \tr[\N [M_{\x_k} ] ~\Lambda [ \rho^{(k)}  ] ]  \label{eq:dualrelation}
\eea
in terms of noise maps on a state $\Lambda$ and a detector $\N$. The maps can be generally identified by channel and detector tomography, which are however not feasible in a NISQ environment.


The description can be illustrated with a unitary noise as follows. Suppose that a detector contains the type of noise that flips a measurement outcome, $M_{\x_k} \mapsto X M_{\x_k} X$ with a Pauli matrix $X$. Or, a state suffers the same type of noise $\rho^{(k)} \mapsto X \rho^{(k)} X$ while a detector remains noiseless as $M_{\x_k}$. The probabilities are given as follows, respectively,
\bea
\tr[ M_{\x_k}    X \rho^{(k)}  X],~\tr[ X M_{\x_k} X \rho^{(k)}  ].  \label{eq:p1}  
\eea
Finally, both a state and a detector suffer noise, $\rho^{(k)} \mapsto \sqrt{X} \rho^{(k)} \sqrt{X}$  and $M_{\x_k} \mapsto \sqrt{X} M_{\x_k} \sqrt{X}$, in which a detection event appears with a probability in the following
\bea
\tr[ \sqrt{X} M_{\x_k}  \sqrt{X}  \sqrt{X} \rho^{(k)}  \sqrt{X}] \label{eq:p2}
\eea
All of the probabilities in Eqs (\ref{eq:p1}) and (\ref{eq:p2}) in the above are identical. Given the statistics of measurement outcomes only, it is not possible to analyze the source of noise that have appeared in a measurement.

 \subsubsection*{Mapping measurement readout to state discrimination} 
 
We here provide an operational interpretation of a measurement in a quantum algorithm. It is supposed a complete measurement is performed, throughout. That is, no additional POVM element apart from the two-outcome POVM in Eq. (\ref{eq:comm}) exists in the description of a detection event. 

\begin{itemize}
\item Assumption 1. A two-outcome measurement on each of the single qubits in a quantum algorithm is complete. 
\end{itemize}

As an example, let us begin with the Deutsch algorithm \cite{deutsch}. Recall that the Deutsch problem is to determine if a function $f : \{0,1 \} \rightarrow \{ 0,1\}$ is constant $(f(0)=f(1))$ or balanced $(f(0) \neq f(1))$. Suppose that the {\it a priori} probabilities are given by $1/2$ for each. The Deutsch algorithm is constructed such that an input state $|0\rangle|A\rangle$ is transformed to a resulting state $|f(0) +f(1)\rangle |A\rangle$ by a single use of the quantum realization of the function $U_f : |\x\rangle |\y\rangle\mapsto |\x\rangle |\y\oplus f(\x)\rangle $.

\begin{enumerate}
\item Preparation  

$ |0\rangle |1\rangle_A$, where $A$ denotes an ancilla qubit\\

\item Unitary transformations : 

i) $H \otimes H:  |0\rangle  |1\rangle_A\mapsto ( \frac{ | 0\rangle + |1\rangle}{\sqrt{2}} ) \otimes (\frac{ | 0\rangle - |1\rangle}{\sqrt{2}})_A $

ii) $U_f : $\\ 

$ ( \frac{ | 0\rangle + |1\rangle}{\sqrt{2}} ) \otimes (\frac{ | 0\rangle - |1\rangle}{\sqrt{2}})_A $ \\

$ \mapsto \frac{1}{\sqrt{2}} (  | 0\rangle + (-1)^{f(0) + f(1)} |1\rangle ) \otimes (\frac{ | 0\rangle - |1\rangle}{\sqrt{2}})_A $    \\

iii) $H \otimes \id : $\\ 

$\frac{1}{\sqrt{2}} (  | 0\rangle + (-1)^{f(0) + f(1)} |1\rangle ) \otimes (\frac{ | 0\rangle - |1\rangle}{\sqrt{2}})_A $    \\

$ \mapsto |f(0)\oplus f(1)\rangle \otimes (\frac{ | 0\rangle - |1\rangle}{\sqrt{2}})_A $    \\

\item Measurement :  $|f(0)\oplus f(1)\rangle  \rangle \mapsto  f(0)\oplus f(1)  $.
\end{enumerate}
Let $U_{\mathrm{Deutsch}}$ denote the overall circuit of the Deutsch algorithm, where calls $U_f$ only once. One can find that the dynamics works as the transformation in the following, 
\bea
U_{\mathrm{Deutsch}} ~:~  |0\rangle|A\rangle \mapsto |f(0) +f(1)\rangle |A\rangle \nonumber
\eea
where $f(0)+f(1)$ is either $0$ or $1$. A measurement in the computational basis is designed for the optimal detection of the solution bits, which is achieved by optimal quantum state discrimination \cite{s1,s2,s3, s4,s5,s6, s7,s8,s9,eldar}. Note that the illustration can be straightforwardly extended to the Deutsch-Jozsa algorithm \cite{DJ}.

A measurement in a quantum algorithm, i.e., individual measurement in the computation basis on single qubits, can be interpreted as an optimal detection of a desired outcome that give rise to a solution in an algorithm. This is achieved by optimal discrimination of qubit states in each register, for the resulting states after a quantum circuit, see also the example of the Deutsch algorithm in the above. In other words, a quantum algorithm should be designed such that individual measurements on single qubits in the computational basis fulfills the task. 

This can be formalized as follows. From Eq. (\ref{eq:nstate}), the $k$-th qubit state $\rho_{\y}^{(k)} = \tr_{\bar{k}}  | \varphi_{\y} \rangle_{1,\cdots, n,A} \langle  \varphi_{\y} |$. Note that the label $\{\y\}$ has been written to indicate a set of possible states according to a designed quantum algorithm. Note also that the {\it a priori} probabilities denoted by $\{ q_{\y}\}$ for those states $\{\rho_{\y}^{(k)} \}$ are provided by a designed quantum algorithm. This defines the problem of optimal state discrimination in the ensemble $\{q_{\y}, \rho_{\y}^{(k)} \}$. Then, a measurement in the computational basis performs optimal discrimination. Note that it is a two-outcome measurement.

\begin{figure}[h]
\begin{center}	
	\includegraphics[width=0.48
\textwidth]{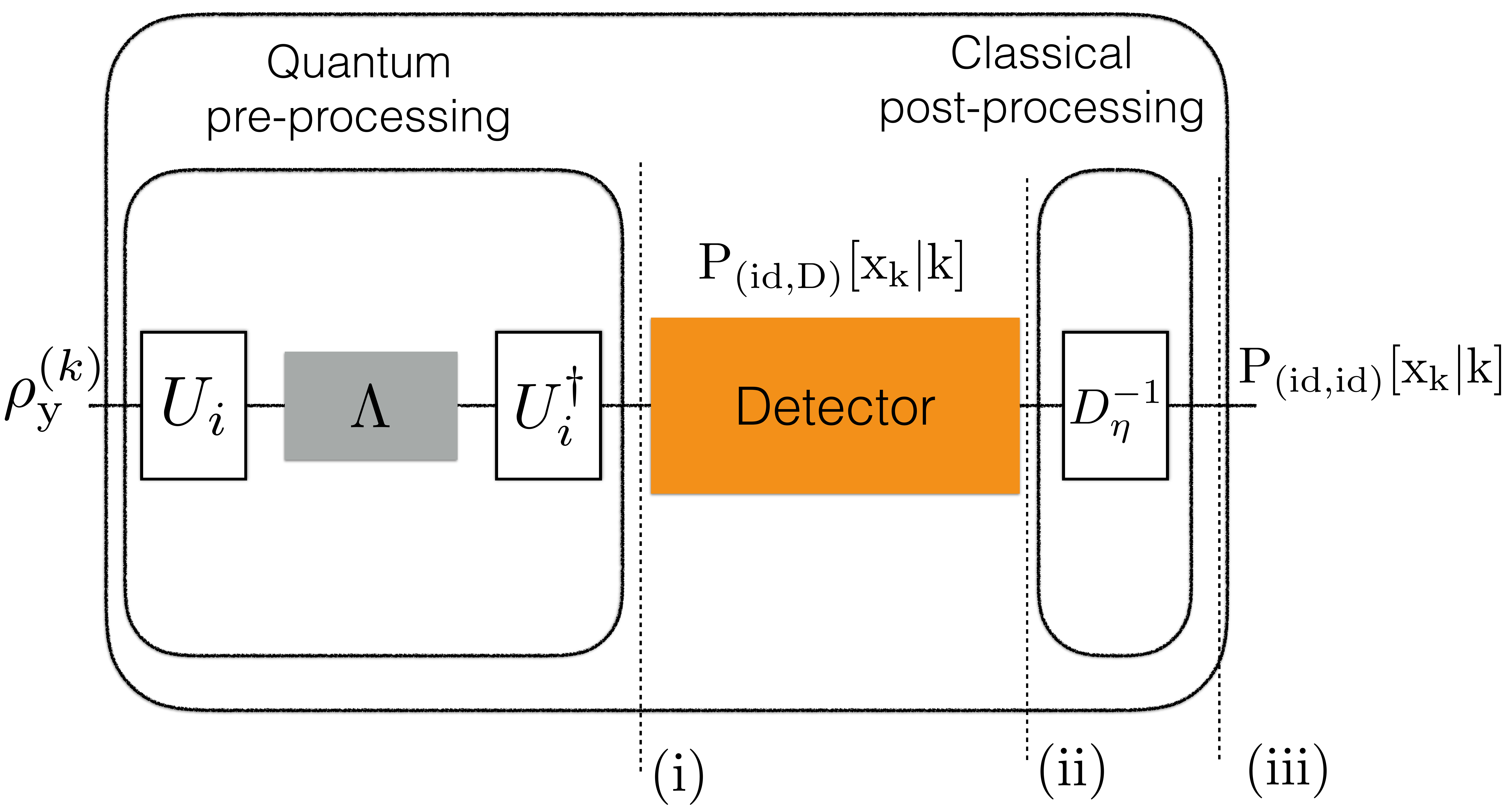}
	\caption{ The scheme for mitigating errors in measurement readout is proposed. (i) Quantum pre-processing performs twirling a unknown channel $\Lambda$ by applying unitaries before a measurement, see Eq. (\ref{eq:3twirl}). This aims to transform a unknown channel to a depolarization map. (ii) Detection events are obtained after depolarizing noise on qubit states. The statistics is found as $\mathrm{P_{(\i,D)} [x_k|k]}$. (iii) Since the channel-twirl is applied, it suffices to deal with depolarization noise only. The inverse map $D_{\eta}^{-1}$ denotes the classical post-processing in Eq. (\ref{eq:correction}) that makes corrections according to the noise rate $\eta$.}
	\label{figscheme}
\end{center}	
\end{figure}

\subsection{Mitigation of noise on quantum detectors}

So far, we have described the sources of noise in measurement readout, and also signified that a measurement in the computational basis performs two-outcome optimal state discrimination for single-qubit states. Having the characterizations on noise and a measurement, we now proceed to devising the scheme of error mitigation in measurement readout. 

The strategy is to depolarize noise on states in order to preserve the optimality of the computational-basis measurement and also, by fixing the type of noise as depolarization, ultimately to perform the classical post-processing efficiently. To clarify the effectiveness of the strategy, in noise model in Eq. (\ref{eq:dualrelation}), we assume that $\N\approx \i$. That is, we restrict to the case that the measurement errors are mostly due to noise on a state right before a detector whereas a detector itself works almost as desired. 
\begin{itemize}
\item Assumption 2. Errors in measurement readout are dominated by noise existing in individual qubits right before a detector that works almost as desired. 
\end{itemize}
In addition, we also assume that the effect of crosstalk in a measurement of multiple qubits is much smaller than noise on individual states. That is, measurement errors are dominated by noise on individual states.  
\begin{itemize}
\item Assumption 3. The effect of crosstalk in measurement readout is nigligible. 
\end{itemize}
Or, the effect of crosstalk in multiple detectors is ignored at the time that we devise the error mitigation scheme, which might also work later even in the presence of such an effect.

Then, from Assumption 2, let us consider an ensemble of single-qubit states $\{q_{\y}, \Lambda[ \rho_{\y}^{(k)} ] \}$ right before a detector where the channel $\Lambda$ is unknown. This means that, due to the unknown channel, a measurement prepared in the computational basis does not perform optimal discrimination. An optimal POVM should be updated according to the ensemble $\{q_{\y}, \Lambda[ \rho_{\y}^{(k)} ] \}$.  However, this is not feasible in practice, since the update requires channel tomography for individual qubits and moreover the type of noise may vary in time in a realistic scenario. 

It turns out that a one-parameter depolarization map in the following,
\bea
D_{\eta} [\rho] = (1-\eta) \rho + \eta   \id /2,  \label{eq:dep}
\eea
preserves an optimal two-outcome measurement in quantum state discrimination \cite{bae2019, bae2020}. As soon as a depolarization map is found as noise on states, a measurement in the computational basis perform optimal state discrimination. It has been known that the channel-twirl can transform a unknown channel $\Lambda$ to a depolarization map: 
\bea 
\Lambda \mapsto \int d\mu (U) U^\dagger \Lambda [ U \rho U^\dagger] U = D_{\eta_{\Lambda}} [\rho ] \label{eq:twirling}
\eea
where the average is performed over the Haar measure, the uniform measure in the space of unitary operators, and $\eta_{\Lambda}$ is determined by a channel $\Lambda$. It is worth to emphasize that the channel-twirl can be implemented by local operations and classical communication only, where ancillary systems are not needed. In practice, a few number of unitary transformations can be applied in the channel-twirl \cite{gross, bae2019, cirac}. The quantum pre-processing is the step to apply a single-qubit channel-twirl right before a detection event.  

The advantages of performing the channel-twirl before a detection event are twofold. First of all, as it is shown with Assumption 2, a measurement in the computational basis remains as an optimal one. It has also been demonstrated that optimal discrimination can be enhanced by preserving an optimal measurement \cite{bae2019}. Next, the channel-twirl fixes the type of noise appearing on states and then consequently in detection events. It is straightforward to deduce the relation between probability distributions for measurement readout of the $k$-th qubit,
\bea
\mathrm{P}_{(\i, D_{\eta_{\Lambda}}) }[\x_k| k ] = (1-\eta_{\Lambda}) \mathrm{P}_{(\i,\i) }[\x_k | k] + \frac{\eta_{\Lambda}}{2}.\label{eq:reln}
\eea
The only unknown parameter in the above is $\eta_{\Lambda}$ that can be determined by a unknown channel $\Lambda$. With an estimate to the parameter, it is possible to correct the statistics of measurement outcomes. The classical post-processing then transforms $\mathrm{P}_{D}[\x_k | k ]$ to $\mathrm{P}_{\i}[\x_k | k]$.

\subsubsection*{  Quantum pre-processing for individual qubits}

We also make a further assumption on noise on qubit states, that the noisy channel is well approximated by a Pauli channel as follows,
\bea
\Lambda_0 [\cdot] = p_0 \i (\cdot) + p_{\x} X(\cdot) X+ p_{\y} Y(\cdot) Y+ p_{\mathrm{z}} Z(\cdot) Z \label{eq:pch}
\eea
with $p_{i}\geq 0$ and $\sum_{i}p_i =1$ where $i=0,\x,\y,\mathrm{z}$. A good approximation means that we have $\| \Lambda -\Lambda_0 \|_{\mathrm{cb}} <\epsilon$ for sufficiently small $\epsilon > 0$, where $\| \cdot \|_{\mathrm{cb}} $ denotes the norm of complete boundness \cite{kitaev, cb}. The probability of distinguishing two channels $\Lambda$ and $\Lambda_0$ is sufficiently small. 
\begin{itemize}
\item Assumption 4. Noise on individual qubits is well approximated by a Pauli channel.
\end{itemize}
Then, for Pauli channels for qubit states, the channel-twirl can be performed by three unitaries as follows,
\bea
\Lambda_0 \mapsto D_{\eta_{\Lambda} } [\rho] = \frac{1}{3}\sum_{ G \in \{ U,V,W \}}  G^\dagger \Lambda_0 [ G \rho G^\dagger] G,\label{eq:3twirl}
\eea
where $U= \id$, 
\bea
V = \frac{1}{2} \left[ \begin{array}{ccc} 1-i & -1-i \\ 1-i  & 1+ i  \end{array} \right]~ \mathrm{and} ~W = \frac{1}{2} \left[ \begin{array}{ccc} -1 - i & -1 - i \\ 1 - i  & - 1 + i  \end{array} \right]. \nonumber 
\eea
Note that the set of three unitaries is not unique, and more collections can be found in Ref. \cite{bae2019,bae2020}. Note also the relation that $\eta_{\Lambda} = ( p_{\x} + p_{\y} + p_{\mathrm{z}})/4$, which is yet unknown since a noisy channel $\Lambda$ is not identified. The quantum pre-processing aims to transform a unknown noisy channel to a depolarization map. 
 
\subsubsection*{ Quantum pre-processing for multiple qubits}

In a quantum algorithm, measurement readout is performed on multiple qubits. Noise on those qubits right before a measurement is generally described by an $n$-qubit map on a large-size Hilbert space $\H^{\otimes n}$. The channel-twirl involves a number of unitary transformations over $n$ qubits in order to depolarize an $n$-qubit map. From Assumption 3 that specifies no effect of crosstalk in multiple detectors, measurements on individual qubits are independent. If the goal is to preserve the optimality of the computational-basis measurement, it suffices to depolarize individual qubits only but not an $n$-qubit map. This leads to a huge simplification in the number of unitary transformations in the quantum pre-processing. 

We introduce a collective channel-twirl with three unitary transformation in Eq. (\ref{eq:3twirl}) for an $n$-qubit map $\Lambda_{0}^{(n)}$ as follows,
\bea
&& \mathcal{M}^{(n)} [\rho_{1\cdots n}]  = \nonumber  \\
&& \frac{1}{3}\sum_{ G \in \{ U,V,W\} }  (G^{\otimes n})^\dagger \Lambda_{0}^{(n)} [ G^{\otimes n} \rho_{1\cdots n} (G^{\otimes n})^\dagger] G^{\otimes n},~~~~~ \label{eq:3twirlm}
\eea
for a multipartite qubit states $\rho_{1\cdots n} \in S(\H\otimes\cdots \otimes \H)$. Although the resulting map $\mathcal{M}^{(n)}$ is not a depolarization map in general, its reduction to a single qubit corresponds to a depolarization map, 
\bea
D_{\eta_{\Lambda} } [\rho_k] =  \tr_{\bar{k}} \mathcal{M}^{(n)} [\rho_{1\cdots n}].  \nonumber
\eea
Thus, it is shown that the collective channel-twirl perform channel twirling an individual channel. In doing so, a measurement in the computational basis for individual qubits remains optimal. We summarize that the collective channel-twirling in Eq. (\ref{eq:3twirlm}) can be used to preserve an optimal measurement on individual qubits. 

\subsubsection*{Classical post-processing}

By fixing the type of noise on qubit states as depolarization, we recall the relation of probabilities in Eq. (\ref{eq:reln}). For convenience, for the $k$-th qubit let $\mathrm{P}_{(\i,D)}[ \x_k | k ] = 1/2 + \alpha$ with $\alpha\in[-1/2, 1/2]$. The reciprocal relation is obtained as follows,
\bea
\mathrm{P}_{(\i,\i) }[\x_k | k ]  & = & \frac{1}{(1-\eta_{\Lambda})} \left(\mathrm{P}_{ (\i,D) }[\x_k | k ] - \frac{\eta_{\Lambda}}{2} \right) \nonumber \\
& = & \frac{1}{2} + \frac{\alpha}{1-\eta_{\Lambda}}. \label{eq:correction}
\eea
Note that for $\alpha<0$ we have $\mathrm{P}_{(\i,\i)} [\x_k | k ] < \mathrm{P}_{ (\i, D) }[\x_k | k ]$ and for $\alpha \geq 0$, $\mathrm{P}_{ (\i,\i) } [\x_k | k ] \geq \mathrm{P}_{ (\i, D) }[x| k ]$. Given an estimated value $\eta_{\Lambda}$, it is possible to update the statistics of measurement outcomes from $\mathrm{P}_{(\i, D) }[\x_k | k ]$ to $\mathrm{P}_{( \i,\i)}[\x_k | k ]$. The scheme of classical post-processing has also been proposed with quantum detector tomography \cite{osm}, up to the uncertainty in the type of noise.

\subsubsection*{ Summary of the scheme and the overhead}

The proposed method of mitigating measurement errors is devised under the aforementioned assumptions. When applying it to quantum algorithms, the steps are summarized as follows. 
\begin{enumerate}
\item Quantum pre-processing: after a circuit and before a measurement, the collective channel-twirl is applied with a set of three single-qubit unitaries, see Eq. (\ref{eq:3twirlm}). 
\item Classical post-processing: after detection events, the statistics of measurement outcomes is updated according to $\eta_{\Lambda}$, see Eq. (\ref{eq:correction})
\end{enumerate}
In practical applications, the value $\eta_{\Lambda}$ can be obtained from specifications of NISQ devices. For IBM quantum devices, the specifications are often reported, e.g. $2$-$10\%$. When such information is not available {\it a priori}, an error rate in measurement readout can be estimated, which is cost effective compared to detector tomography. 
 
In the quantum pre-processing, it is required to perform twirling a channel over multiple qubits. We note that a unitary $2$-design for channels over multiple qubits contains a number of unitaries, where the number increases exponentially. We have devised the collective channel-twirl in Eq. (\ref{eq:3twirlm}) that contains three unitaries only. One can also refer it as G-channel-twirl since the task is associated with a set $G$ of three unitaries.  

The collective channel-twirl leads to a depolarization channel in a single-qubit level if a reduced map $ \tr_{\bar{k}}[\Lambda_{0}^{(n)}[\cdot] ] $ is a Pauli channel. That is, we have the relation in the following, see also Eq. (\ref{eq:3twirlm})
\bea
&& \tr_{\bar{k}}~~ \frac{1}{3}\sum_{ G \in \{ U,V,W\} }  (G^{\otimes n})^\dagger \Lambda_{0}^{(n)} [ G^{\otimes n} \rho_{1\cdots n} (G^{\otimes n})^\dagger] G^{\otimes n} \nonumber \\
&=& (1- \eta_{0}) \rho_k + \eta_{0}   \id/2 \nonumber
\eea
where $\rho_k$ is the $k$-th qubit state, $\rho_k = \tr_{\bar{k}} \rho_{1\cdots n}$ and $\eta_0 = \eta_{ \tr_{\bar{k}} [\Lambda_{0}^{(n)}]}$. We have used the notation that $\tr_{\bar{k}}$ is the trace over all qubits but the $k$-th one and $\tr_{\bar{k}} [\Lambda_{0}^{(n)} ] $ defines a local map on the $k$-th qubit state. 

The advantage of the quantum pre-processing is that three unitary gates are only required regardless to the number of qubits. The set of three unitaries immediately implements a depolarization channel in a single-qubit level. It is required to apply local unitaries coherently, i.e., $G^{\otimes n}$ for $G\in \{ U,V,W\}$, which is feasible with NISQ devices as single-qubit operations can be realized with a high precision. In doing so, the measurement in the computational basis remains an optimal one in a NISQ environment. In Sec. \ref{IBM}, the pre-processing is realized in two quantum algorithms to mitigate the errors in measurement readout.

\begin{figure*}[h]
\begin{center}	
	\includegraphics[width=0.9\textwidth]{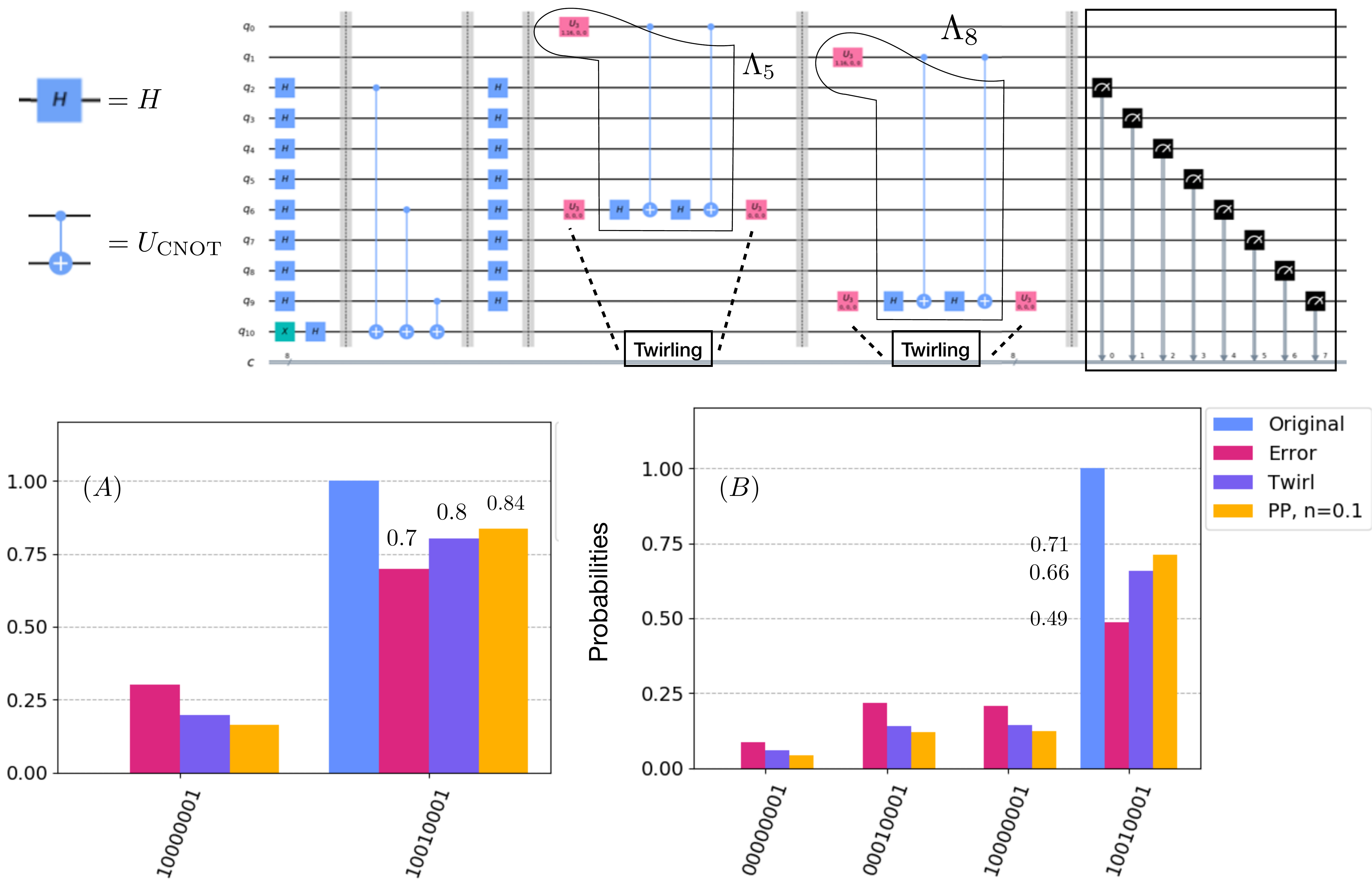}
	\caption{ 
	A quantum circuit of the BV algorithm that finds the solution is $\vec{s} = 10010001$ is shown. In the circuit, the first two qubits are ancillar qubits to be used to introduce noise in the $5$th and $8$th system qubits, see Fig. \ref{noisemodel}. Three CNOT gates in the $1$-st, the $5$-th, and the $8$-th registers implement the quantum oracle in Eq. (\ref{eq:qoracle}). Two boxes characterized by quantum channels $\Lambda_5$ and $\Lambda_8$ introduce noise on the $5$-th and the $8$-th qubits, respectively. The maps introduce noise in detection events in the $5$-th and $8$-th qubits. The quantum pre-processing applies the channel-twirl in Eq. (\ref{eq:3twirl}) before and after a noisy channel. The gray box after a measurement denotes the classical post-processing. The BV algorithm is ideal when neither $\Lambda_5$ nor $\Lambda_8$ is applied. $(A)$ The statistics of measurement outcomes is shown when detection events in the $5$-th register contain noise, i.e., with $\Lambda_5$ only. If no noise exists in the BV algorithm, it is with the unit the probability to obtain the bit-string $\vec{s} = 10010001$ (blue). When noise exists in the $5$-th detector, i.e., with $\Lambda_5$, the probability drops to $0.7$ (red). With the quantum pre-processing, the probability increases to $0.8$ (violet). Together with the classical post-processing, the probability increases up to $0.84$ (orange). $(B)$ The statistics of measurement outcomes are shown when detection events in the $5$- and $8$-qubits both contain noise. The probability of obtaining the bit-string $\vec{s} = 10010001$ in a measurement drops to $0.49$ by noise on two detectors, and increases to $0.66$ by the channel-twirl and then up to $0.71$ together with the classical post-processing. 
	}	\label{BVsimulation}
\end{center}	
\end{figure*}

\section{ Numerical Simulation }
\label{demonstration}

In this section, we apply the proposed method of error mitigation to quantum algorithms and demonstrate their performances under the condition that the addressed four assumptions are fulfilled. The BV and QAE algorithms are particularly considered. The BV algorithm is based on oracle queries and has applications in Learning Parities with Noise \cite{smolin}. The QAE algorithm contains the building blocks common to other ones, such as QAA and QFT that are key elements in quantum search, quantum period-finding, and quantum factoring. Although the QAE algorithm can be simplified \cite{simplified, rudy}, we here consider the original algorithm that contains the building blocks: it is more useful to find if the proposed scheme may work, or not. Moreover, recently it is found that the QAE algorithm can be applied to quantum finance problems \cite{finance}. 

The numerical simulation is performed under the aforementioned four assumptions. From Assumption 1, a measurement on single qubits is complete. From Assumption 2 and Assumption 3, POVMs describing detectors are noiseless whereas qubit states right before detection events contain noise. From Assumption 4, noise on qubit states is described by Pauli channels. Then, the quantum pre- and the classical post-processings are applied.

\begin{figure*}[h]
\begin{center}	
\includegraphics[width=0.85\textwidth]{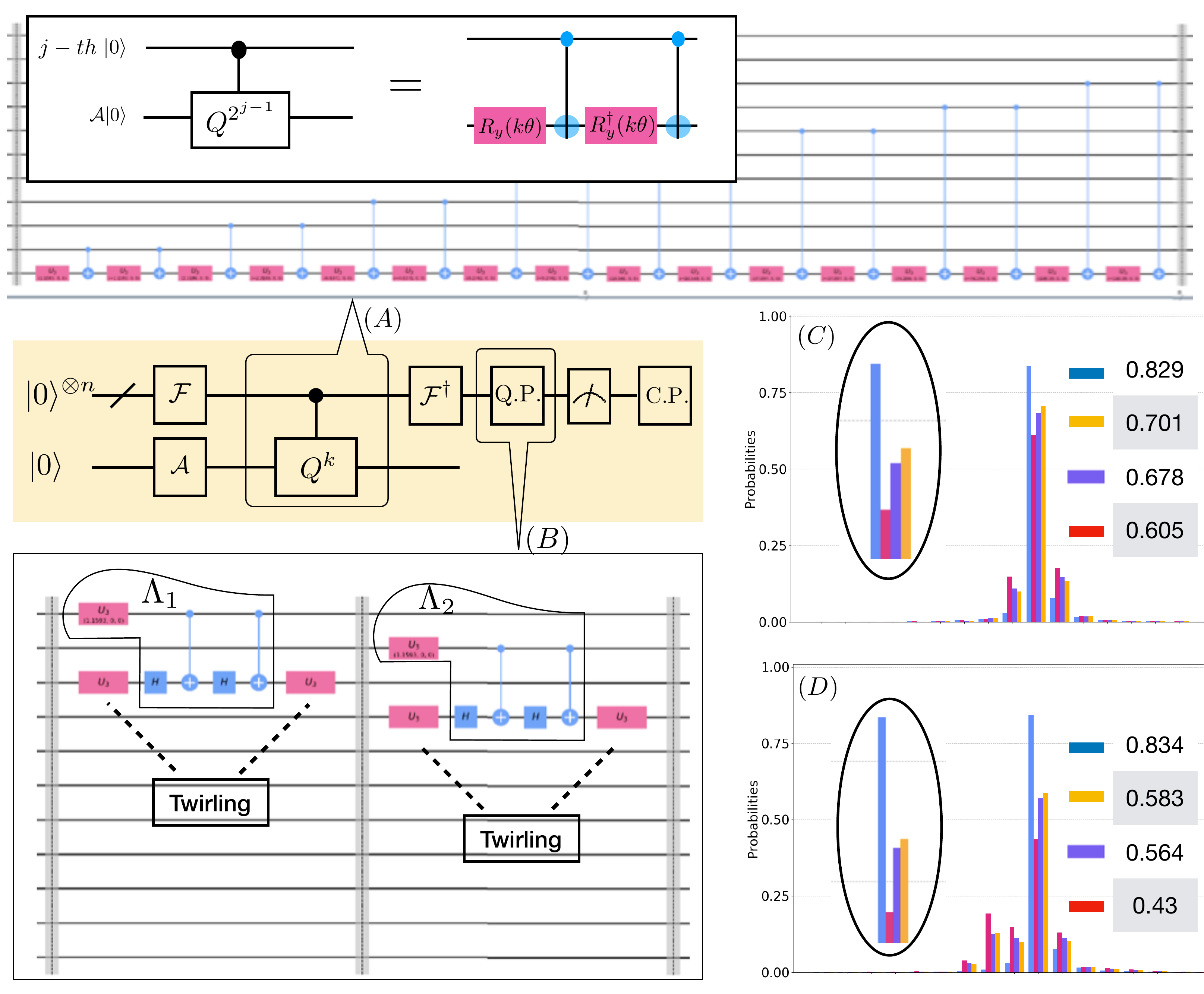}
	\caption{The dynamics in the QAE algorithm is composed of three steps, i) QFT, ii) QAA, and iii) inverse QFT. Note that an ancilla qubit is prepared as $\mathcal{A} |0\rangle  = \sqrt{1-p} |0\rangle + \sqrt{p}|1\rangle$. In the simulation with $8$ qubits, we have put $p=0.3$, and then the QAE algorithm aims to estimate the value $p$ from measurement outcomes. In $(C)$ and $(D)$, the maximum peaks are centered at $\tilde{p}=0.2974$. The quantum pre-processing is applied before a measurement, after which the classical post-processing is performed with measurement outcomes. (A) The QAA algorithm is structured by a series of $Q^k$ for $k=2^{j-1}$ where $j=1,\cdots,n$. Recall that $\theta_0 = 2\sin^{-1} (\sqrt{p})$, see also Eq. (\ref{eq:est}), from which we have $\theta_0 \approx 1.1593$. It also holds that $\mathcal{A} = R_y(\theta_0)$ and $Q = R_y(2\theta_0)$. The gate $Q^{k}$ can be implemented by two rotations, $R_y(k\theta )$ and its inverse $R_{y}^{\dagger}(k\theta )$, with two CNOT gates. (B) It is supposed that qubits in the first or both the first and the second registers suffer noise $\Lambda_Y [\cdot] = 0.7\i (\cdot)+ 0.3 Y (\cdot)Y$, see Fig. \ref{noisemodel}. In the quantum pre-processing, the channel-twirl is performed with three unitaries, see Eq. (\ref{eq:3twirl}). (C) The peaks are centered at $\tilde{p} = 0.2974$ close to $p=0.3$. In the ideal case (blue), the probability of having $\tilde{p} = 0.2974$ is $0.829$. By a probabilistic $Y$-noise in the first qubit, described by $\Lambda_1$, the probability gets down to $0.605$ (red). By the quantum pre-processing only, it increases up to $0.678$ (violet). Finally, classical post-processing makes it up to $0.701$ (orange) with $\eta=0.1$. (D) When the first and both detectors are noisy by $\Lambda_1$ and $\Lambda_2$, the collective channel-twirl is applied as the quantum pre-processing, and then the classical post-processing are applied collectively on the both registers. It is shown that mitigation of noise in detectors can enhance the statistics of measurement outcomes.} \label{fig:qae}
\end{center}	
\end{figure*}

The simulation is performed in a Qiskit simulator. In the Qiskit package, single-qubit operations are programmed in the following forms, 
\bea
U_2 (\phi,\lambda) =  \left[ \begin{array}{ccc} 1/\sqrt{2} &  -e^{i\lambda} /\sqrt{2}  \\  e^{i \phi} /\sqrt{2}   &  e^{i\lambda +i\phi} /\sqrt{2}   \end{array}  \right], \nonumber 
\eea
and
\bea
U_3 (\theta, \phi, \lambda) =   \left[ \begin{array}{ccc} \cos (\theta/2) & -  e^{i\lambda} \sin(\theta /2) \\   e^{i \phi} \sin(\theta /2)  &   e^{i\lambda + i\phi} \cos(\theta /2)  \end{array} \right]. \nonumber
\eea
For instance, the Hadamard transform can be realized by $H= U_2 (0,\pi)$ and the rotation in Eq. (\ref{eq:ry}) can be found as $R_y(\theta) = U_3 (\theta,0,0)$. 

With the gates, the circuits for the BV and the QAE algorithms are shown in Figs. \ref{BVsimulation} and \ref{fig:qae}. The first two registers are for ancilla qubits that implement noisy channels on two systems qubits right. The noisy channels are realized as it is shown in Fig. \ref{noisemodel}. The circuits are then with $8$ system qubits. The results of a proof-of-principle demonstration are also shown in Figs. \ref{BVsimulation} and \ref{fig:qae}. For both BV and QAE algorithms, it is demonstrated that the statistics of measurement outcomes is enhanced by the proposed method. 

The results of numerical simulation show that when the assumptions are valid, the proposed scheme of error mitigation can efficiently work to enhance the statistics of measurement outcomes. We remark that the results also demonstrate the usefulness of the collective channel-twirl in Eq. (\ref{eq:3twirlm}). This is closely related to Assumption 3 that the effect of crosstalk is much smaller compared to noise on individual qubits. One can also consider the channel-twirl on individual qubits an incoherent manner, in which it is found that an improvement is not presented even if a depolarization channel on individual qubits has been obtained. In the following, let us briefly summarize the BV and the QAE algorithms.

\subsection{ The Berstein-Vazirani algorithm }

Let us considers a function $f:\{ 0,1\}^{n} \rightarrow \{0,1 \}$ that works as follows,  
\bea
f:  \vec{\x} \mapsto \sum_{i=1}^n \x_{i}s_i, ~~\mathrm{for~all}~\vec{x} \in \{0,1 \}^{n} \label{eq:BV}
\eea
and for some $\vec{s} \in \{0,1 \}^{n}$. The BV problem is then to find the bit-string $\vec{s}$. In the BV algorithm to solve the problem, a unitary transformation in the following is implemented as a quantum realization of the map in Eq. (\ref{eq:BV}),
\bea
U_f : |\vec{\x}\rangle |y\rangle \mapsto  | \vec{\x} \rangle | y\oplus f(\vec{\x})\rangle \label{eq:qoracle}
\eea
where $\oplus$ denotes bit-wise addition. The BV algorithm has shown that a single application of the unitary transformation $U_f$ can find the string $\vec{s}$. The steps are summarized as follows.
\begin{enumerate}
\item Preparation 
$ |0\rangle ^{\otimes n } |1\rangle_A$, where $A$ denotes an ancilla qubit
\item Unitary transformations : 
i) $H^{\otimes n }\otimes H :$
$|0\rangle ^{\otimes n } |1\rangle_A\mapsto ( \frac{ | 0\rangle + |1\rangle}{\sqrt{2}} )^{\otimes n }\otimes (\frac{ | 0\rangle - |1\rangle}{\sqrt{2}})_A $\\
ii) $U_f : $ 
$( \frac{ | 0\rangle + |1\rangle}{\sqrt{2}} )^{\otimes n } \otimes (\frac{ | 0\rangle - |1\rangle}{\sqrt{2}})_A  $
$\mapsto \frac{1}{\sqrt{2^n}} \sum_{\vec{x}} (-1)^{f(\vec{\x})} |\vec{x}\rangle \otimes (\frac{ | 0\rangle - |1\rangle}{\sqrt{2}})_A $ 
iii) $H^{\otimes n} \otimes \id : $
$ \frac{1}{\sqrt{2^n}} \sum_{\vec{x}} (-1)^{f(\vec{\x})} |\vec{x}\rangle \otimes (\frac{ | 0\rangle - |1\rangle}{\sqrt{2}})_A$ 
$\mapsto | \vec{s} \rangle \otimes (\frac{ | 0\rangle - |1\rangle}{\sqrt{2}})_A $ 
\item Measurement : $\{ M_0, M_1\}^{\otimes n}$
\end{enumerate}
The measurement corresponds to the map from the resulting state to the solution, $|\vec{s} \rangle \mapsto \vec{s}$. A quantum circuit for the BV algorithm with $8$ qubits is shown in Fig. \ref{BVsimulation}

\begin{figure*}[h]
\begin{center}	
	\includegraphics[width=1.0\textwidth]{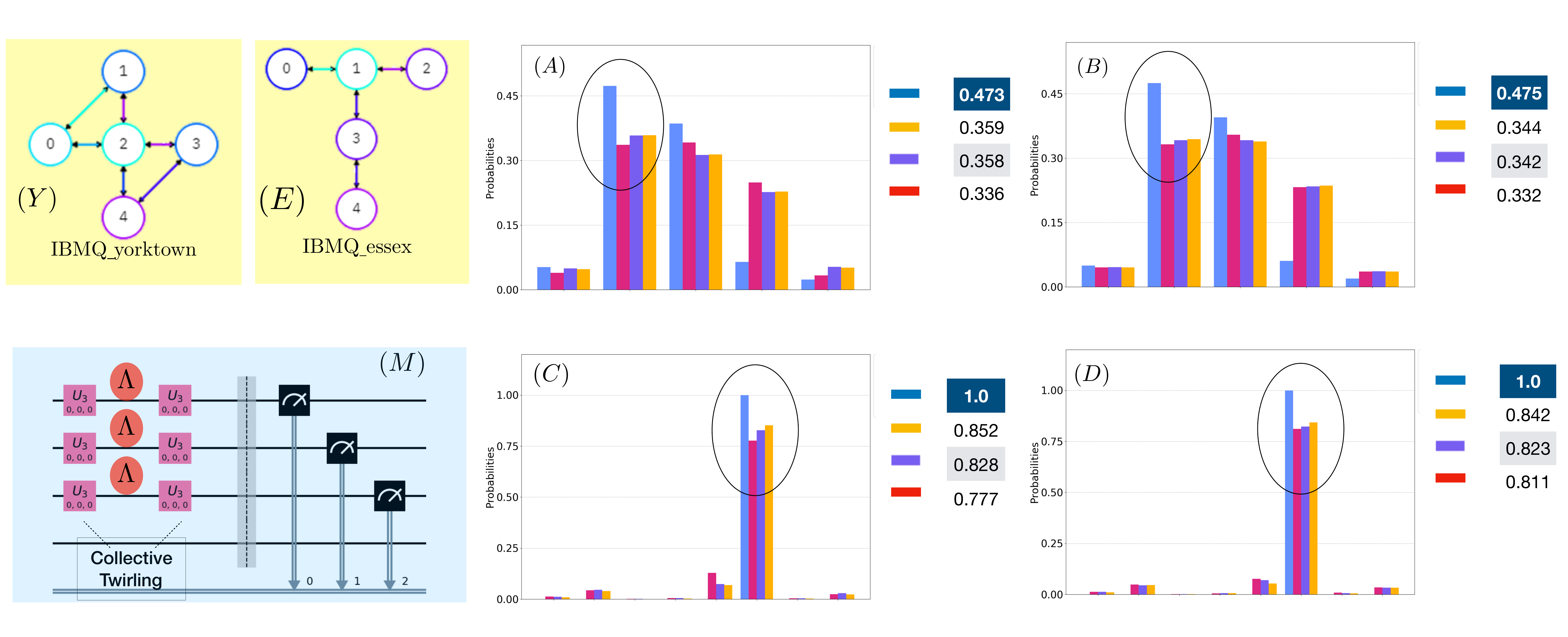} 
	\caption{ The QAE and BV algorithms are realized in IBM quantum computers. The algorithms are performed with $3$ system qubits and $1$ ancilla qubit. (Y) The topology of $5$ qubits in $\mathrm{IBMQ\underbar{~}yorktown}$ is shown, in which the QAE algorithm is realized. (E) The topology of $5$ qubits in $\mathrm{IBMQ\underbar{~}yorktown}$ is shown, in which the BV algorithm is realized. (M) In the quantum pre-processing, the collective twirl in Eq. (\ref{eq:3twirlm}) is performed. Noisy channels that may exist as $\Lambda$ between the two unitaries are then transformed to depolarization maps. The classical post-processing is applied to each of the three registers. (A) The $3$-qubit QAE algorithm is realized in $\mathrm{IBMQ\underbar{~}yorktown}$ with error mitigation in the three registers. The parameter $\eta=2\%$ is considered in the classical post-processing. The peak is at $\tilde{p} =0.1464$ for $p=0.3$. (B) The $3$-qubit QAE algorithm is realized in $\mathrm{IBMQ\underbar{~}yorktown}$ with measurement errors in the first register only. The parameter $\eta=5\%$ is considered in the classical post-processing. (C) The $3$-qubit BV algorithm is realized in $\mathrm{IBMQ\underbar{~}essex}$ with error mitigation in the three registers. The parameter $\eta=2\%$ is considered in the classical post-processing. The solution in the algorithm is given as $\vec{s}=101$. (D) The $3$-qubit BV algorithm is realized in $\mathrm{IBMQ\underbar{~}essex}$ with error mitigation in the $1$st detector only. The parameter $\eta=5\%$ is considered in the classical post-processing. All results demonstrate that the proposed scheme of error mitigation can enhance the statistics of measurement outcomes in the IBM Q devices. }
	\label{IBMresult}
	\end{center}	
\end{figure*}

\subsection{ Quantum amplitude estimation algorithm  }

Suppose that unitary transformation denoted by $\mathcal{A}$ is given as follows, 
\bea
\mathcal{A} : |0\rangle^{\otimes n} \mapsto  |\Psi\rangle = \sqrt{1- p }  | \Psi_0 \rangle +   \sqrt{p} |\Psi_1\rangle. \label{eq:qaepro}
\eea
where $| \Psi_0 \rangle$  and $| \Psi_1\rangle $ are normalized states and $p$ is real. The problem is to find the amplitude $p$, that the QAE algorithm aims to estimate. Note that one can define $\theta_p$ that satisfies $\sin^2 \theta_p = p$. 

In the QAE algorithm, two building blocks are the QFT on $m$ qubits,
\bea
\mathcal{F}_m : | x\rangle \mapsto  \frac{1}{\sqrt{2^m}} \sum_{  y=0}^{2^m-1} \exp[\frac{2\pi i}{2^m} xy] |y\rangle. \nonumber
\eea
and an operator from QAA is implemented as follows,
\bea
Q = (\id - 2 |0\rangle \langle 0|^{\otimes n} ) (\id - 2| \Psi_0\rangle \langle \Psi_0 |). \nonumber
\eea
The unitary transformation $\mathcal{F}_{m}^{\dagger}$ denotes the inverse QFT. Note that the operator $Q$ is defined in the two-dimensional space spanned by $| \Psi_0 \rangle$  and $| \Psi_1\rangle $. In fact, the eigenvalues of the operator can be found by $\lambda_{\pm} = \exp[{\pm i 2\theta_p}]$ and the eigenvectors $|\Psi_{\pm}\rangle = (|\Psi_1 \rangle \pm i |\Psi_0\rangle  )/\sqrt{2}$, respectively. In the QAE algorithm, the operator $Q$ is applied conditioned on $m$ qubits, $U_Q : |j\rangle |\y\rangle \mapsto |j\rangle Q^j |\y\rangle$. 

Then, the QAE algorithm can be summarized as follows. 

\begin{enumerate}

\item Preparation  \\

$ |0\rangle ^{\otimes m } |0\rangle_{A}^{\otimes n }$, where $A$ denotes ancilla qubits. \\

\item Unitary transformations: \\

i) $\mathcal{F}_m \otimes \mathcal{A} : $ \\

$ |0\rangle ^{\otimes m } |0\rangle_{A}^{\otimes n} \mapsto  \frac{1}{\sqrt{2^m}} \sum_{  y=0}^{2^m-1}  |y\rangle  \otimes |\Psi\rangle_A $\\

ii) $ U_Q : $ \\

$\frac{1}{\sqrt{2^m}} \sum_{  y=0}^{2^m-1}  |y\rangle  \otimes |\Psi\rangle_A \mapsto $ \\

$ |\Phi\rangle = \frac{1}{\sqrt{2^m}} \sum_{  y=0}^{2^m-1}  e^{i2\theta_p y} | y \rangle  \otimes \frac{i}{\sqrt{2}} e^{i\theta_p}|\Psi_+ \rangle_A $  \\

$ - \frac{1}{\sqrt{2^m}}\sum_{  y=0}^{2^m-1}  e^{ - i2\theta_p y} | y \rangle  \otimes \frac{i}{\sqrt{2}} e^{ - i\theta_p}|\Psi_- \rangle_A $    \\

iii) $\mathcal{F}_{m}^{\dagger}  \otimes \id : $\\

$ |\Phi\rangle \mapsto$\\

$ | 2^m \frac{\theta_p}{\pi} \rangle \otimes  \frac{i}{\sqrt{2}} e^{i\theta_p } |\Psi_+\rangle_A + |2^m (1-\frac{\theta_p}{\pi})\rangle\otimes  \frac{i}{\sqrt{2}} e^{i\theta_p } |\Psi_-\rangle_A   $  \\

\item Measurement : $\{ M_0, M_1\}^{\otimes m} $\\
\end{enumerate}
The measurement reads outcomes $z_1 z_2 \cdots z_m$ in binary numbers, which are converted to a decimal one $z \in \{0,\cdots, 2^m-1 \}$. This provides an estimator 
\bea
\tilde{p} = \sin^2( z \pi/2^m) \label{eq:est}
\eea
an approximation to the amplitude $p$. It has been shown that the estimator satisfies $| p  - \tilde{p} | \leq O( 2^{-m} )$ with a probability greater than $8/\pi^2$. Thus, we have that $\tilde{p}$ converges to $p$ with a high probability as $m$ tends to be a large number. 

A quantum circuit for the QAE algorithm with $8$ qubits is shown in Fig. \ref{fig:qae}. The $8$-qubit circuit finds $\tilde{p}=0.2974$ for $p=0.3$ encoded by $\mathcal{A}$.

\section{ Applications to IBM quantum computers}
\label{IBM}

The proposed scheme of error mitigation in a quantum measurement is applied to IBM Q devices. The goal is find if the proposed scheme works to mitigate measurement errors in IBM Q devices. Note that measurement errors in IBM quantum devices are within the range $2$-$10\%$ whereas single-qubit gates are with errors around $0.1\%$. One may expect the possibility that single-qubit gates in the proposed scheme together with the classical post-processing would be used to mitigate measurement errors. 

To make it clear when one can expect the proposed scheme is effective, we briefly summarize the four assumptions addressed when the scheme is devised. From Assumption 1, a measurement on single qubits is complete. In Assumption 2, it is supposed that errors in measurement readout is dominated by noise on a state right before a detector, but not a detector itself. Hence, the quantum pre-processing aims to manipulate a noisy channel on states. Assumption 3 considers that the effect of crosstalk in detection events is much smaller than other errors. Hence, the collective channel-twirl in Eq. (\ref{eq:3twirlm}) can be applied to twirling single-qubit channels. Then, Assumption 4 asserts that noise on qubits right before a detector is well approximated by a Pauli channel. Under the assumptions, it is confirmed by numerical simulation in a Qiskit simulator, that the proposed scheme can mitigate errors in measurement readout. 


We have implemented the BV and QAE algorithms with $3$ qubits and $1$ ancilla qubit in IBM Q devices. The BV algorithm is realized in $\mathrm{IBM\underbar{~}essex}$ and the QAE algorithm is in $\mathrm{IBM\underbar{~}yorktown}$. The circuits are composed by consider the topology of connected qubits in the devices, see Fig. \ref{IBMresult}. In both cases, the collective channel-twirl and the classical post-processing are applied to all of the three system qubits. The results in Fig. \ref{IBMresult} obtained in Feb. 2020 show that the proposed scheme of error mitigation in measurement readout works to enhance the statistics of measurement outcomes. The statistics is obtained by collecting the data of $8192$ shots.

The details of the quantum pre- and classical post-processing are as follows. The quantum pre-processing is placed as the final step of a quantum circuit or right before detectors. This aims to depolarize a noisy channel that potentially exists in between final gates and detectors. In case no noise exists in between, the channel-twirl corresponds to the identity map, that causes an error up to $0.2\%$ in IBM Q devices due to two single-qubit gates. This is yet much lower than that of measurement errors. In the classical post-processing shown in Eq. (\ref{eq:correction}), the parameter $\eta_{\Lambda}$ is needed while a noisy channel $\Lambda$ is not identified. From the reported data \cite{data}, we have taken $\eta=2\%$ or $\eta=5\%$ in the classical post-processing. 

\section{Conclusion}
\label{remark}

We have presented a method of mitigating errors in measurement readout in quantum algorithms. The scheme applies quantum pre- and classical post-processing,. The pre-processing singles out depolarization as the type of noise on a qubit state before a measurement. Tomography of noise on a qubit state is thus circumvented. Assuming that a detector is almost noise-free, the classical post-processing on measurement outcome can be fixed as an inverse map of a depolarization channel. 

The scheme is carefully devised by clarifying the assumptions made on a measurement in quantum algorithms and quantum noise. With Assumption 1, an operational meaning of a measurement on single qubits in quantum algorithms can be found as optimal state discrimination to obtain a desired bit-string in measurement readout. By Assumption 2, it is asserted that noise for measurement errors can be manipulated by the channel-twirl with single-qubit gates only, such that the computational-basis measurement remains for optimal detection. With Assumption 3 that the effect of crosstalk is negligible, the collective channel-twirl is devised to keep the optimality of measurements on single qubits. Under Assumption 4, it is shown that the quantum pre-processing of the channel-twirl with three unitaries only can perform channel twirling for unknown type of noise on single-qubit states.  

The proposed scheme works to enhance the statistics of the measurement outcomes in quantum algorithms. Numerical simulation with $8$-qubit circuits is provided for the BV and the QAE algorithms, which contain QFT and QAA as building blocks. We have then applied the scheme to IBM Q devices to mitigate measurement errors. Improvements in the statistics of measurement outcomes are presented. Our scheme may be used as a microarchitecture of measurement readout with NISQ information processing. We envisage that the scheme can be applied to NISQ devices in practice. 

We remark that the results presented with IBM Q devices does not imply that the four assumptions on measurement readout are fulfilled in the devices. The assumptions aim to identify the theoretical model of a measurement, the sources of quantum noise, and errors of measurement readout, in order to make it clear to list the possible noise on real devices. In fact, it has been found that crosstalk in multiple detectors is evident \cite{crosstalk}, that rules out Assumption 3 immediately. 

We point out the role of the collective channel-twirl in the relation of the four assumptions and the performance of error mitigation in measurement readout. In fact, additional quantum gates can be used to mitigate the effect of crosstalk \cite{ion} \cite{12}. In future investigations, it is worth to consider other quantum circuits to analyze the relations among the address assumptions, the collective channel-twirl, and the performance of the proposed scheme for error mitigation. It may allow us to learn detailed properties of NISQ devices, e.g.,  sources of noise in measurement readout or the effect of crosstalk in multiple detectors. It is also desired to characterize the usefulness of the collective channel-twirl in the presence of crosstalk in multiple detectors. Finally, it is left an open question to find how the hierarchy of computational complexity is ultimately related to enhancement by error mitigation on NISQ devices \cite{nature}.

\section*{Acknowledgement}

We acknowledge use of the IBM Q for this work. The views expressed are those of the authors and do not reflect the official policy or position of IBM or the IBM Q team. This work is supported by National Research Foundation of Korea (NRF-2019M3E4A1080001, $\mathrm{O_2 N}$-$\mathrm{Q_2 A}$), an Institute of Information and Communications Technology Promotion (IITP) grant funded by the Korean government (MSIP), ITRC Program (IITP2020-2018-0-01402).


\begin{thebibliography}{99}





\bibitem{shor} P. W. Shor, "Algorithms for quantum computation: discrete logarithms and factoring," {\it Proceedings 35th Annual Symposium on Foundations of Computer Science}, Santa Fe, NM, USA, 1994, pp. 124-134.

\bibitem{simon} D. R. Simon, "On the power of quantum computation," {\it Proceedings 35th Annual Symposium on Foundations of Computer Science}, Santa Fe, NM, USA, 1994, pp. 116-123.

\bibitem{grover} L. K. Grover, Quantum Mechanics helps in searching for a needle in a haystack, Phys. Rev. Lett. {\bf 79} 325, 1997.

\bibitem{bv}  E. Bernstein and U. Vazirani "Quantum Complexity Theory". {\it SIAM Journal on Computing}. 26 (5): 1411–1473, 1997.

\bibitem{deutsch} D. Deutsch, "Quantum Theory, the Church-Turing Principle and the Universal Quantum Computer". {\it Proceedings of the Royal Society of London A.} {\bf 400}, 97–117, 1985. 

\bibitem{ambainis} A. Ambainis, Understanding Quantum Algorithms via Query Complexity, arxiv:1712.06349.

\bibitem{book} M. A. Nielsen and I. Chuang, Quantum Computation and Quantum Information, Cambridge University Press, 2000.

\bibitem{qaa} G. Brassard, P. Høyer, M. Mosca, A. Tapp, "Quantum Amplitude Amplification and Estimation". arXiv:quant-ph/0005055.

\bibitem{preskill} J. Preskill, Quantum Computing in the NISQ era and beyond, arXiv:1801.00862. 

\bibitem{google} F. Arute {\it et. al.}, "Quantum supremacy using a programmable superconducting processor", Nature {\bf 574}, 505–510, 2019. 

\bibitem{konig} S. Bravyi, D. Gosset,  and R. Koenig, "Quantum advantage with shallow circuits" Science Vol. 362, Issue 6412, pp. 308-311, 2018.

\bibitem{vqe} A. Peruzzo, J. McClean, P. Shadbolt, M.-H. Yung, X.-Q. Zhou, P. J. Love, A. Aspuru-Guzik, and J. L. O’Brien, "A variational eigenvalue solver on a photonic quantum processor", Nature Communications 5, 5213, 2014.

\bibitem{qaoa} E. Farhi, J. Goldstone, and S. Gutmann, “A quantum approximate optimization algorithm”, arXiv:1411.4028. 


\bibitem{qec}  P. W. Shor, "Scheme for reducing decoherence in quantum computer memory". Physical Review A. {\bf 52} (4): R2493–R2496, 1995. 

\bibitem{data} Bibek Pokharel, Namit Anand, Benjamin Fortman, Daniel Lidar, "Demonstration of fidelity improvement using dynamical decoupling with superconducting qubits", Phys. Rev. Lett. {\bf121}, 220502, 2018.

\bibitem{ion} S. Debnath {\it et. al.}, "Demonstration of a small programmable quantum computer with atomic qubits
", Nature {\bf 536}, 63-66, 2016. 

\bibitem{crosstalk} Y. Chen, M. Farahzad, S. Yoo, and T.-C. Wei, "Detector Tomography on IBM 5-qubit Quantum Computers and Mitigation of Imperfect Measurement", Phys. Rev. A {\bf 100}, 052315, 2019.
	

\bibitem{allnisq} C. Song, J. Cui, H. Wang, J. Hao, H. Feng and Y. Li, "Quantum computation with universal error mitigation on a superconducting quantum processor", Science Advances, Vol. 5, no. 9, 5686, 2019. 

\bibitem{n1} S. Endo, S. C. Benjamin, and Y. Li, "Practical Quantum Error Mitigation for Near-Future Applications", Phys. Rev. X {\bf 8}, 031027, 2018.

\bibitem{n2} K. Temme, S. Bravyi, J. M. Gambetta, "Error mitigation for short-depth quantum circuits", Phys. Rev. Lett. {\bf119}, 180509, 2017.

\bibitem{osm} F. B. Maciejewski, Z. Zimborás, and M. Oszmaniec, "Mitigation of readout noise in near-term quantum devices
by classical post-processing based on detector tomography", arXiv:1907.08518. 

\bibitem{gtech} S. S. Tannu and M. K. Qureshi, "Mitigating Measurement Errors in Quantum Computers by Exploiting State-Dependent Bias" MICRO '52: Proceedings of the 52nd Annual IEEE/ACM International Symposium on Microarchitecture, 279, 2019.


\bibitem{solovay} C. M. Dawson and M. A. Nielsen, "The Solovay-Kitaev algorithm", arXiv:quant-ph/0505030.


\bibitem{s1} A. S. Holevo, "Remarks on optimal quantum measurements”, Problemy Peredachi Informatsii, {\bf 104} 51–55, 1974.

\bibitem{s2} H. P. Yuen, R. S. Kennedy and M. Lax, "Optimum Testing of Multiple Hypotheses in Quantum Detection Theory
", {\it IEEE Transactions on Information Theory} 21 125, 1975.

\bibitem{s3} C. W. Helstrom, Quantum detection and estimation theory, vol 84 (Academic press New York), 1976.

\bibitem{s4} A. Chefles, "Quantum state discrimination", Contemporary Physics 41 401, 2000.

\bibitem{s5} J. A. Bergou, U. Herzog and M. Hillery, Quantum state estimation (Springer) pp 417-465, 2004.

\bibitem{s6} J. A. Bergou, "Quantum state discrimination and selected applications", J. Phys.: Conf. Ser. 84 012001, 2007.

\bibitem{s7} J. A. Bergou, "Discrimination of quantum states", J. Mod. Opt. 57 160, 2010.

\bibitem{s8} S. M. Barnett and S. Croke, "Quantum state discrimination", Adv. Opt. and Photon. 1 238, 2009.

\bibitem{s9} J. Bae and L.-C. Kwek, "Quantum state discrimination and its applications", J. Phys. A: Math. Theor. 48 083001, 2015. 

\bibitem{DJ} D. Deutsch and R. Jozsa, "Rapid solutions of problems by quantum computation". {\it Proceedings of the Royal Society of London A.} 439: 553–558, 1992. 

\bibitem{eldar} Y. C. Eldar, A. Megretski and G. C. Verghese, "Designing optimal quantum detectors via semidefinite programming," in {\it IEEE Transactions on Information Theory}, vol. 49, no. 4, pp. 1007-1012, 2003.

\bibitem{bae2019} S. Kechrimparis, T. Singal, C.M. Kropf, and J. Bae,  "Preserving Measurements for Optimal State Discrimination over Quantum Channels", Phys. Rev. A {\bf 99} (6), 062302, 2019.

\bibitem{bae2020} S. Kechrimparis, C.M. Kropf, F. Wudarski, and J. Bae, "Channel Coding of a Quantum Measurement", arXiv:1908.10735, to appear in IEEE Journal on Selected Areas in Communications, 2020.

\bibitem{gross}D. Gross, K. Audenaert, and J. Eisert, "Evenly distributed unitaries: on the structure of unitary designs", J. Math. Phys., {\bf 48}, 052104, 2007.

\bibitem{cirac} W. Dür, J. I. Cirac, M. Lewenstein, and D. Bruß, "Distillability and partial transposition in bipartite systems" Phys. Rev. A {\bf 61}, 062313, 2000.


\bibitem{cb} V. Paulsen, "Completely Bounded Maps and Operator Algebras", Cambridge Press, 2002.

\bibitem{kitaev}  A. Kitaev, "Quantum computations: algorithms and error correction", Russ. Math. Surv. {\bf 52}, 1191, 1997.

\bibitem{smolin} A. W. Cross, G. Smith, and J. A. Smolin, "Quantum learning robust against noise
", Phys. Rev. A {\bf 92}, 012327 (2015).

\bibitem{simplified} S. Aaronson and P. Rall, "Quantum Approximate Counting, Simplified", arXiv:1908.10846.

\bibitem{rudy}  Y. Suzuki {\it et. al.}, "Amplitude estimation without phase estimation", Quantum Information Processing {\bf 19}, 75,  2020. 

\bibitem{finance} S. Woerner and D. J. Egger, "Quantum Risk Analysis", npj Quantum Information {\bf5}, 15, 2019. 

\bibitem{nature} A. Kandala {\it et. al.,}, "Error mitigation extends the computational reach of a noisy quantum processor", Nature vol. 567, 491–495, 2019.

\bibitem{12} M. Gong {\it et. al.}, "Genuine 12-qubit entanglement on a superconducting quantum processor", Phys. Rev. Lett. {\bf 122}, 110501, 2019. 





\end{thebibliography}
\end{document}